\documentclass[aps,floats,nofootinbib,amssymb,
preprint,superscriptaddress]{revtex4}
\usepackage{graphicx}

\begin{document}


\title{Charged Higgs phenomenology \\ in the flipped two Higgs doublet model}

\author{Heather~E.~Logan}
\email{logan@physics.carleton.ca}
\affiliation{Ottawa-Carleton Institute for Physics, Carleton University,
Ottawa K1S 5B6 Canada}

\author{Deanna MacLennan}
\affiliation{Ottawa-Carleton Institute for Physics, Carleton University,
Ottawa K1S 5B6 Canada}

\date{\today}

\begin{abstract}
We study the phenomenology of the charged Higgs boson in the
``flipped'' two Higgs doublet model, in which one doublet gives mass
to up-type quarks and charged leptons and the other gives mass to
down-type quarks.  We present the charged Higgs branching ratios and
summarize the indirect constraints.  We extrapolate existing LEP
searches for $H^+H^-$ and Tevatron searches for $t \bar t$ with $t \to
H^+ b$ into the flipped model and extract constraints on $M_{H^+}$ and
the parameter $\tan\beta$.  We finish by reviewing existing LHC
charged Higgs searches and suggest that the LHC reach in this model
could be extended for charged Higgs masses below the $tb$ threshold by
considering $t \bar t$ with $t \to H^+ b$ and $H^+ \to q \bar
q^{\prime}$, as has been used in Tevatron searches.
\end{abstract}

\thispagestyle{empty}
\maketitle


\section{Introduction}

The Standard Model (SM) of electroweak
interactions~\cite{G-S-W:theory} has been stringently tested over the
past twenty years and is in excellent agreement with all collider
data.  The dynamics of electroweak symmetry breaking, however, remain
unknown.  While the simplest possibility is the minimal Higgs
mechanism~\cite{higgs} implemented with a single scalar SU(2) doublet,
many extensions of the SM enlarge the Higgs sector to contain
additional scalars.

Extensions of the SM Higgs sector are tightly constrained by two
pieces of data: (i) the rho parameter, $\rho \equiv M_W^2 / M_Z^2
\cos^2\theta_W \simeq 1$, where $M_W$ ($M_Z$) is the $W^{\pm}$ ($Z$)
boson mass and $\theta_W$ is the weak mixing angle; and (ii) the
absence of large flavor-changing neutral currents.  The first of these
constraints is automatically satisfied by Higgs sectors that contain
only SU(2) doublets (with the possible addition of singlets).  The
simplest such model that contains a charged Higgs boson is a
two-Higgs-doublet model (2HDM).  The second of these constraints is
automatically satisfied by models in which the masses of fermions with
a common electric charge are generated through couplings to exactly
one Higgs doublet; this is known as natural flavor
conservation~\cite{GWP} and prevents the appearance of tree-level
flavor-changing neutral Higgs interactions.\footnote{The 2HDM without
natural flavor conservation is known as the Type III
model~\cite{TypeIII}; for a review of its phenomenology see
Ref.~\cite{Atwood:1996vj}.  In this model the basis chosen for the two
Higgs doublets is somewhat arbitrary; basis-independent methods have
been developed in Refs.~\cite{Davidson:2005cw,Haber:2006ue}.  The
flavor-conserving limit of the Type III model, in which the two Yukawa
matrices for each fermion type are required to be diagonal in the same
fermion mixing basis, has been developed in Refs.~\cite{MFV} under the
name minimal flavor violation.}

Imposing natural flavor conservation, there are four different
ways~\cite{Barnett:1983mm,Barnett:1984zy} to couple the SM fermions to
two Higgs doublets, as summarized in
Table~\ref{tab:Possible-2HDMs}.\footnote{We ignore neutrino masses.}
Each of these four coupling assignments gives rise to a different
phenomenology for the charged Higgs boson $H^{\pm}$.  In each case,
the charged Higgs couplings can be parameterized in terms of the free
parameter $\tan\beta \equiv \langle \Phi_2^0 \rangle / \langle
\Phi_1^0 \rangle$.

\begin{table}
\begin{tabular}{ccccc}
\hline \hline
Model \ & Type I \ & Type II \ & Lepton-specific \ & Flipped \\
\hline
$\Phi_{1}$ & -- & $d,\ell$ & $\ell$ & $d$ \\
$\Phi_{2}$ & $u,d,\ell$ & $u$ & $u,d$ & $u,\ell$ \\
\hline \hline
\end{tabular}
\caption{The four possible assignments of fermion couplings to two
Higgs doublets that satisfy natural flavor conservation.  Here
$u$, $d$, and $\ell$ represent up- and down-type quarks and charged
leptons, respectively.}
\label{tab:Possible-2HDMs}
\end{table}

The Type-I and -II 2HDMs have been studied extensively.  In
particular, the Type-II
2HDM~\cite{Lee:1973iz,Fayet:1974fj,Peccei:1977hh,Fayet:1976cr}, in
which one doublet generates the masses of up-type quarks while the
other generates the masses of down-type quarks and charged leptons,
arises naturally in supersymmetric models; this model has dominated
2HDM collider studies.  The Type-I
2HDM~\cite{Georgi:1978wr,Haber:1978jt}, in which one doublet generates
the masses of all quarks and leptons while the other contributes only
to the $W^{\pm}$ and $Z$ boson masses, has also been widely
considered, particularly in the context of indirect constraints.  More
recently, the lepton-specific
2HDM~\cite{Barnett:1983mm,Barnett:1984zy} has received attention
because of its possible role in neutrino mass~\cite{Aoki:2008av} and
dark matter~\cite{Goh:2009wg} models and its potential to modify the
signatures of the SM-like Higgs at the CERN Large Hadron Collider
(LHC)~\cite{Su:2009fz}.  Charged Higgs phenomenology in this model
has been studied in Refs.~\cite{Aoki:2009ha,Logan:2009uf}.

In this paper we study the remaining model, in which one doublet
$\Phi_u \equiv \Phi_2$ gives mass to up-type quarks and charged
leptons and the other doublet $\Phi_d \equiv \Phi_1$ gives mass to
down-type quarks.  This scenario has been referred to in the
literature as Model IB~\cite{Barnett:1983mm,Barnett:1984zy}, Model
III~\cite{Barger:1989fj}, Model II$^{\prime}$~\cite{Grossman:1994jb},
the flipped 2HDM~\cite{Hmodels}, and the Type-Y
2HDM~\cite{Aoki:2009ha}; following Ref.~\cite{Hmodels} we will call it
the flipped 2HDM.  This coupling structure was first introduced in
Refs.~\cite{Barnett:1983mm,Barnett:1984zy} and its phenomenology has
been studied together with that of the other three models in
Refs.~\cite{Barger:1989fj,Grossman:1994jb,AkeroydLEP,Akeroyd:1998ui,Hmodels,Aoki:2009ha}.
The early studies in Refs.~\cite{Barger:1989fj,Grossman:1994jb}
focused on indirect constraints while Refs.~\cite{AkeroydLEP} and
\cite{Akeroyd:1998ui} considered phenomenology (mainly of the neutral
Higgs bosons) at the CERN Large Electron-Positron (LEP) Collider and
the LHC, respectively.  Reference~\cite{Hmodels} addressed the
coupling patterns of the neutral CP-even Higgs bosons.  Finally,
Ref.~\cite{Aoki:2009ha} presented decay branching ratios for the
charged and neutral Higgs bosons.

Our aim in this paper is to provide a comprehensive study of the
existing constraints and LHC search prospects for the flipped 2HDM.
We begin in Sec.~\ref{sec:model} with an outline of the model
structure of the flipped 2HDM and the relevant Feynman rules for the
charged Higgs couplings.  In Sec.~\ref{sec:brs} we show the decay
branching fractions of the charged Higgs as a function of its mass and
$\tan\beta$ and compare them to those in the Type-II 2HDM.  In
Sec.~\ref{sec:indirect} we review the indirect constraints on the
model, the strongest of which comes from $b \to s \gamma$ and
constrains the charged Higgs mass to be above about 200--300~GeV, the
same as in the Type-II 2HDM.  However, a lighter charged Higgs may be
possible if additional flavor-violating physics leads to cancellations
in the $b \to s \gamma$ rate.  In Sec.~\ref{sec:direct} we thus
reinterpret the existing charged Higgs limits from LEP (due to $e^+e^-
\to H^+H^-$) and the Fermilab Tevatron (due to $t \bar t$ production
with $t \to b H^+$) to provide direct search limits in the flipped
model for the first time.  In Sec.~\ref{sec:lhc} we then review
existing LHC studies of charged Higgs signatures performed in the
context of the Type-II 2HDM and translate them for the first time into
the flipped 2HDM.  We show that while the production cross sections of
the charged Higgs at the LHC are identical to the corresponding modes
in the Type-II 2HDM, the different decay branching fractions below the
$H^+ \to tb$ threshold lead to very different signatures---in
particular, the usual $\tau \nu$ decay of the charged Higgs in the
Type-II model is replaced by decays to quarks ($c \bar b$ and $c \bar
s$) in most of the parameter space.  Above the $tb$ threshold the
decay of the charged Higgs to $tb$ dominates in both the flipped and
Type-II model and existing LHC studies of this channel carry over.
Finally we summarize our conclusions in Sec.~\ref{sec:conclusions}.

\section{The model}
\label{sec:model}

We begin as usual with two complex SU(2)-doublet scalar fields
$\Phi_u$ and $\Phi_d$, 
\begin{equation}
   \Phi_i = \left( \begin{array}{cc} \phi_i^+ \\
     (v_i + \phi_i^{0,r} + i \phi_i^{0,i})/\sqrt{2} \end{array} \right),
   \qquad \qquad i = u, d,
   \label{eq:2HD}
\end{equation}
where the vacuum expectation values (vevs) of the two doublets are
constrained by the $W$ mass to satisfy $v_u^2 + v_d^2 = v_{\rm SM}^2
\simeq 246$~GeV.  To enforce the desired structure of the Yukawa
Lagrangian we impose a discrete symmetry under which $\Phi_d$ and the
right-handed down-type quarks transform according to
\begin{equation}
  \Phi_d \to -\Phi_d, \qquad d_{Rj} \to -d_{Rj},
\end{equation}
while all other fields remain invariant.\footnote{In order to preserve
natural flavor conservation, this discrete symmetry may be broken only
by dimension-two terms in the scalar potential.  We further assume
$CP$ conservation in the scalar potential.}  The Yukawa Lagrangian is
then,
\begin{equation}
   \mathcal{L}_{\rm Yuk} = \sum_{i,j=1}^3
   \left[ y_{ij}^u \bar u_{Ri} \widetilde{\Phi}_u^{\dagger} Q_{Lj}
     + y_{ij}^d \bar d_{Ri} \Phi_d^{\dagger} Q_{Lj}
     + y_{ij}^{\ell} \bar \ell_{Ri} \Phi_u^{\dagger} L_{Lj} 
     \right] + {\rm h.c.},
  \label{eq:Lyuk}
\end{equation}
where $i,j$ are generation indices, $y_{ij}^{u,d,\ell}$ are 3$\times$3
complex Yukawa coupling matrices, $Q_L$ and $L_L$ are the left-handed
quark and lepton doublets, and the conjugate Higgs doublet
$\widetilde{\Phi}$ is given by
\begin{equation}
   \widetilde{\Phi}_i \equiv i \sigma_2 \Phi_i^*
   = \left(\begin{array}{cc} (v_i + \phi_i^{0,r} - i \phi_i^{0,i})/\sqrt{2} \\
     -\phi_i^- \end{array} \right).
\end{equation}

Defining $\tan\beta \equiv v_u/v_d$, the physical charged Higgs boson
is 
\begin{equation}
  H^{\pm} = -\sin\beta \, \phi_d^{\pm} + \cos\beta \, \phi_u^{\pm}.
\end{equation}
The Feynman rules for charged Higgs boson couplings to fermions are 
then given as follows, with all particles incoming:\footnote{For comparison,
the corresponding couplings in the Type-I 2HDM are~\cite{HHG}
\begin{eqnarray} 
   H^+ \bar{u}_i d_j &:& \frac{ig}{\sqrt{2}M_W} V_{ij} 
   \cot\beta \, (m_{ui} P_L - m_{dj} P_R) \nonumber \\
   H^+ \bar{\nu}_i \ell_i &:& -\frac{ig}{\sqrt{2}M_W}
   \cot\beta \, m_{\ell i} P_R,
\label{eq:type1coups}
\end{eqnarray}
with $\tan\beta = v_2/v_1$, where $v_2$ is the vev of the Higgs field
that couples to fermions; the other doublet is decoupled from
fermions.
In the Type-II 2HDM the couplings are~\cite{HHG}
\begin{eqnarray} 
   H^+ \bar{u}_i d_j &:& \frac{ig}{\sqrt{2}M_W} V_{ij}
   (\cot\beta \, m_{ui} P_L + \tan\beta \, m_{dj} P_R) \nonumber \\
   H^+ \bar{\nu}_i \ell_i &:& \frac{ig}{\sqrt{2}M_W} 
   \tan\beta \, m_{\ell i} P_R,
\label{eq:type2coups}
\end{eqnarray}
again with $\tan\beta = v_2/v_1$; this time $v_1$ ($v_2$) is the vev
of the doublet that couples to down-type quarks and charged leptons
(up-type quarks).  
Finally, in the lepton-specific 2HDM the couplings are~\cite{Barger:1989fj}
\begin{eqnarray} 
  H^+ \bar{u}_i d_j &:& \frac{ig}{\sqrt{2}M_W} V_{ij}
  \cot\beta \, (m_{ui} P_L - m_{dj} P_R ), \nonumber \\
  H^+ \bar{\nu}_i \ell_i &:& \frac{ig}{\sqrt{2}M_W}
  \tan\beta \, m_{\ell i} P_R,
\label{eq:leptonspecificcoups}
\end{eqnarray}
with $\tan\beta = v_q/v_{\ell}$, where $v_q$ is the vev of the doublet
that couples to up- and down-type quarks while $v_{\ell}$ is the vev
of the doublet that couples to leptons.}
\begin{eqnarray} 
   H^+ \bar{u}_i d_j &:& \frac{ig}{\sqrt{2}M_W} V_{ij} \,
   (\cot\beta \, m_{ui} P_L + \tan\beta \, m_{dj} P_R) \nonumber \\
   H^+ \bar{\nu}_i \ell_i &:& \frac{ig}{\sqrt{2}M_W}
   \cot\beta \, m_{\ell i} P_R.
\label{eq:flippedcoups}
\end{eqnarray}
Here $V_{ij}$ is the CKM matrix and $P_{L,R} \equiv (1 \mp
\gamma^5)/2$ are the left- and right-handed projection operators.  We
ignore neutrino masses and thus take $\nu_i$ as the flavor eigenstate
corresponding to the charged lepton $\ell_i$.
In particular, the $H^+ \bar u d$ coupling is identical to that in the
familiar Type-II 2HDM (compare Eq.~\ref{eq:type2coups}), while the
$H^+ \bar \nu \ell$ coupling has the opposite $\tan\beta$ dependence.
 
The allowed range of $\tan\beta$ can be constrained by the requirement
that the Yukawa couplings remain perturbative, $y_i^2/4\pi \lesssim
1$.  The top quark Yukawa coupling $y_t = \sqrt{2} m_t / v_{\rm SM}
\sin\beta$ provides a lower limit on $\tan\beta$ and the bottom quark
Yukawa coupling $y_b = \sqrt{2} m_b / v_{\rm SM} \cos\beta$ provides
an upper limit.  Taking $m_t = 171.3$~GeV~\cite{PDG} and $m_b =
4.20$~GeV ($\overline{\rm MS}$ mass~\cite{PDG}), we obtain
\begin{equation}
  0.29 \lesssim \tan\beta \lesssim 150.
\end{equation}
We will work in a range of $\tan\beta$ corresponding to more moderate
values of the Yukawa couplings; for reference, we note that
\begin{equation}
  0.49 \, (0.98) \leq \tan\beta \leq 83 \, (41) 
  \qquad {\rm for} \ y_t, y_b \leq 2 \, (1).
\end{equation}

\section{Charged Higgs boson decays}
\label{sec:brs}

We compute the branching fractions of the charged Higgs in the flipped
2HDM by adapting the public FORTRAN code {\tt HDECAY} version
3.3~\cite{HDECAY}.  {\tt HDECAY} computes the branching fractions and
total width of the SM Higgs and the Higgs bosons of the minimal
supersymmetric standard model (MSSM), including QCD and some
electroweak corrections.  We adapt the charged Higgs part of the code
to the flipped 2HDM by inserting the appropriate $\tan\beta$
dependence for the fermion couplings according to
Eq.~\ref{eq:flippedcoups} and eliminating decays to supersymmetric
particles.  {\tt HDECAY} includes: (i) charged Higgs decays to leptons
at tree level, including final state mass effects; (ii) charged Higgs
decays to quarks including full one-loop QCD corrections and using the
running quark masses in the Yukawa couplings computed to three loops
in QCD and evaluated at the charged Higgs mass; and (iii)
off-shell decays to $t \bar b$ below threshold.  No electroweak or
supersymmetric corrections are included.  Decays via the weak gauge
coupling to $\phi^0 W^+$ (where $\phi^0$ is one of the neutral states
$h^0$, $H^0$, or $A^0$) are generically present; for simplicity we
eliminate these decays by taking $M_{H^0} = M_{A^0} = M_{H^+}$ and
choosing the $h^0$--$H^0$ mixing angle such that the $H^+h^0W^-$
coupling is zero.

\begin{figure}
\resizebox{0.5\textwidth}{!}{\rotatebox{270}{\includegraphics{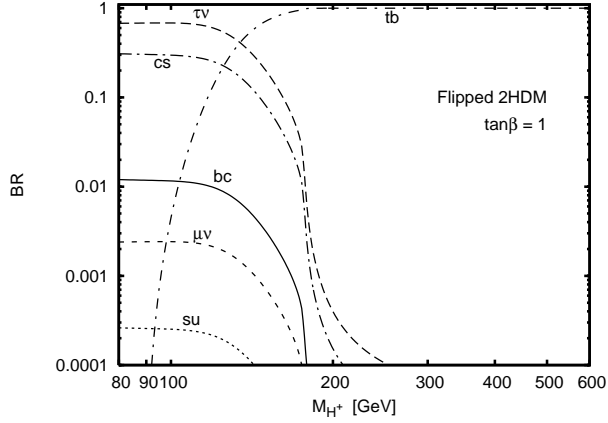}}}
\caption{Charged Higgs branching ratios as a function of $M_{H^+}$
for $\tan\beta=1$ in the flipped 2HDM.  The branching ratios in the
Type-II 2HDM are identical.}
\label{fig:flip1}
\end{figure}

\begin{figure}
\resizebox{\textwidth}{!}{\rotatebox{270}{\includegraphics{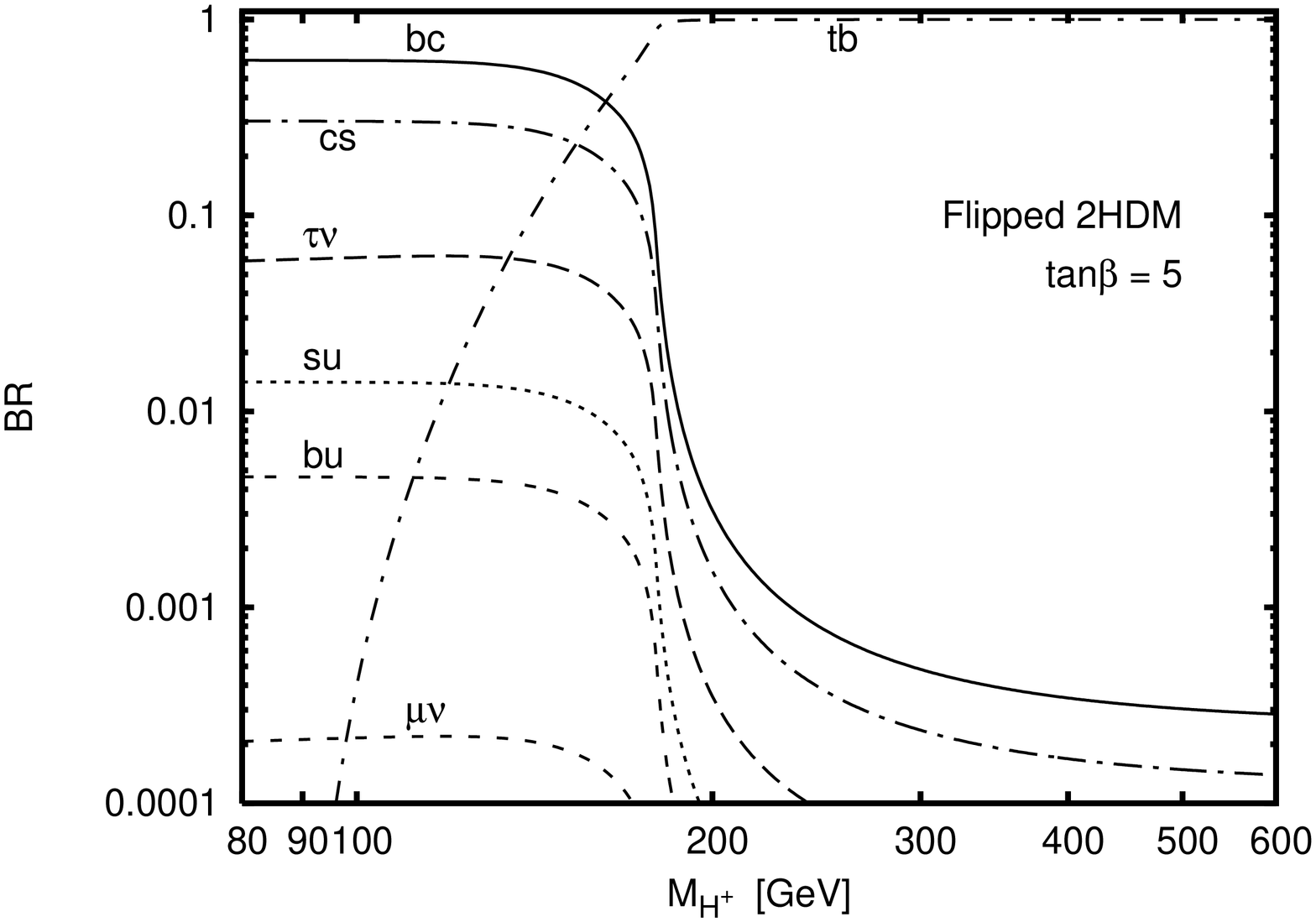}}
\rotatebox{270}{\includegraphics{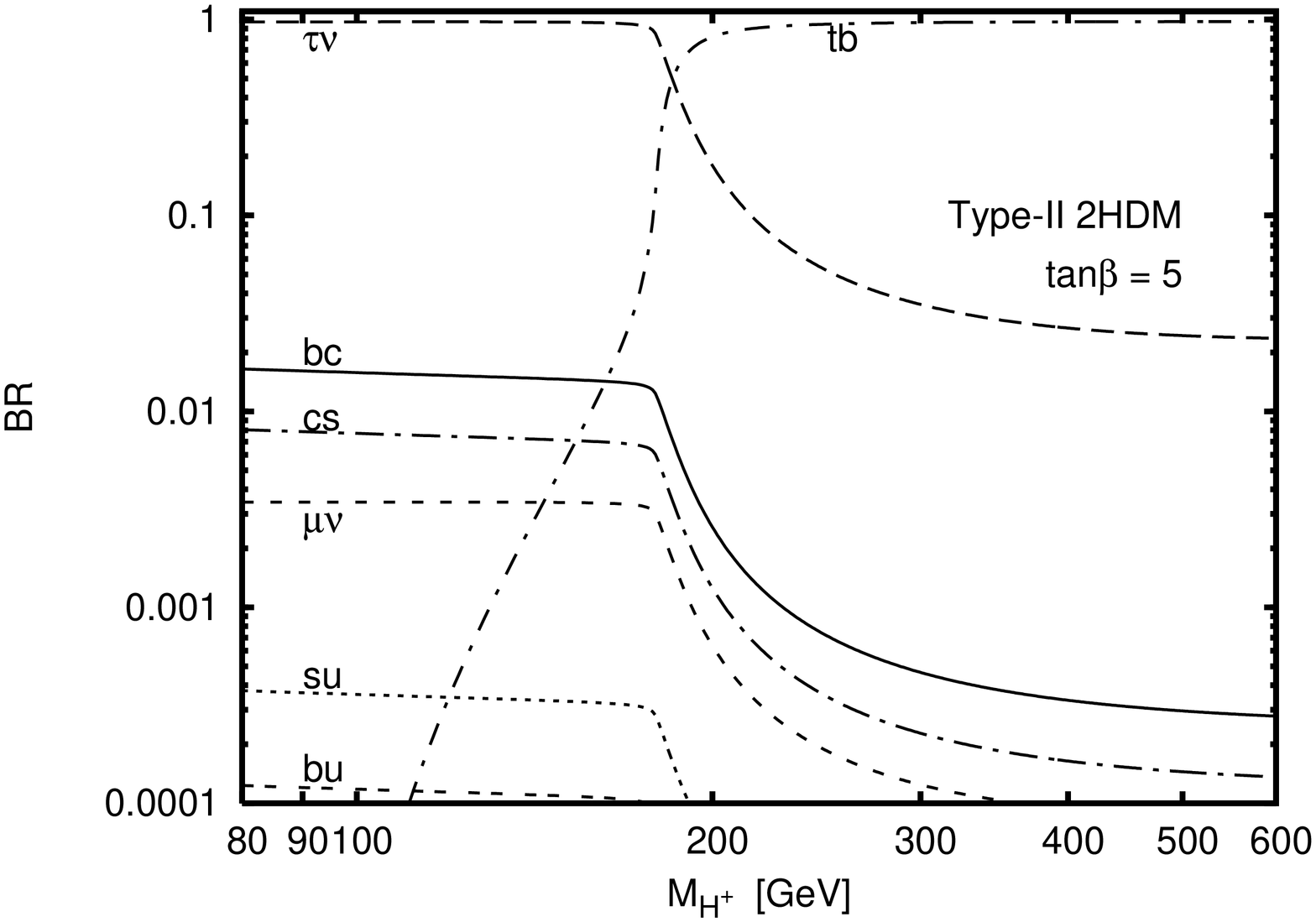}}}
\caption{Charged Higgs branching ratios as a function of $M_{H^+}$
for $\tan\beta=5$ in the flipped (left) and Type-II (right) 2HDMs.}
\label{fig:flip5}
\end{figure}

\begin{figure}
\resizebox{\textwidth}{!}{\rotatebox{270}{\includegraphics{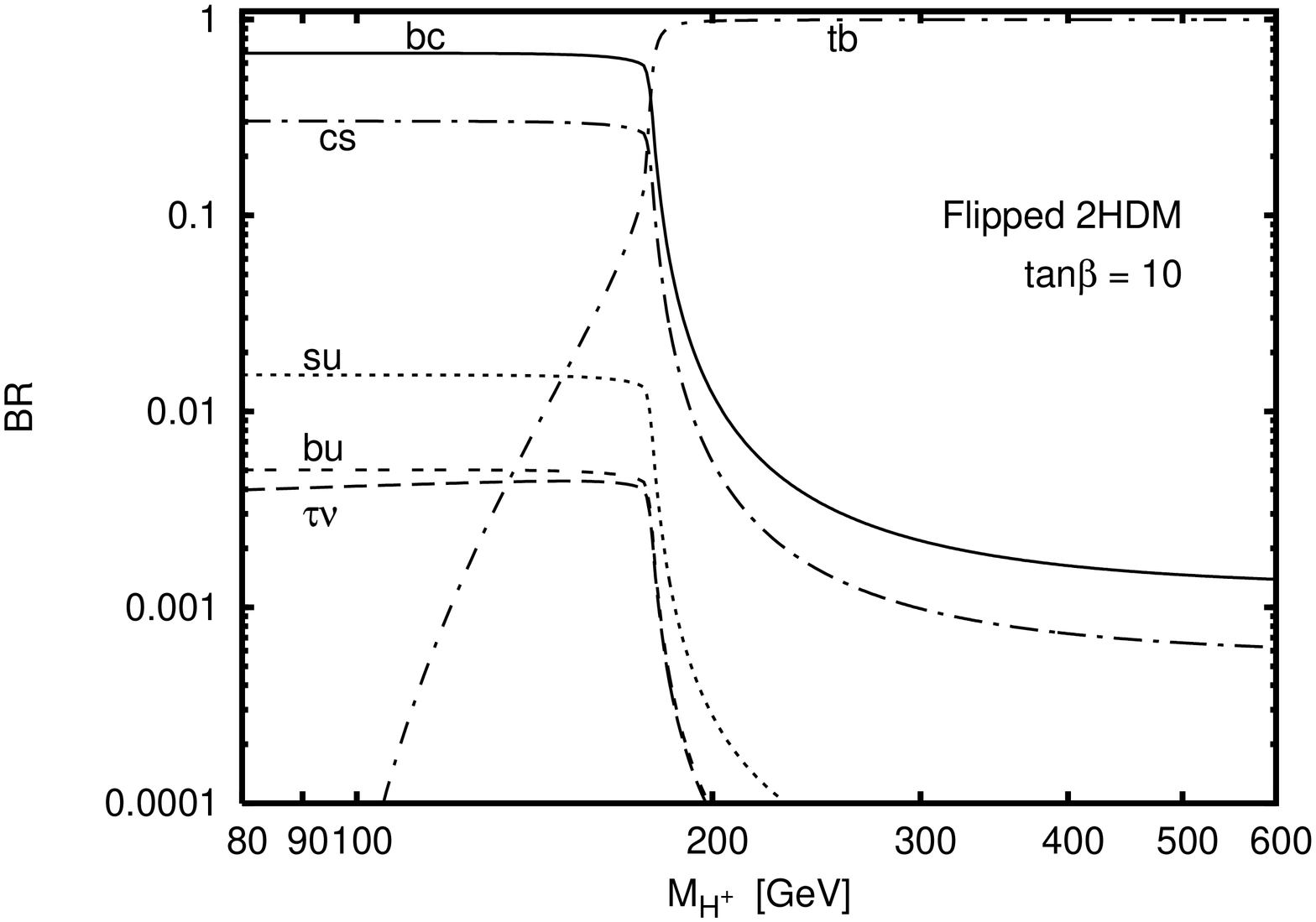}}
\rotatebox{270}{\includegraphics{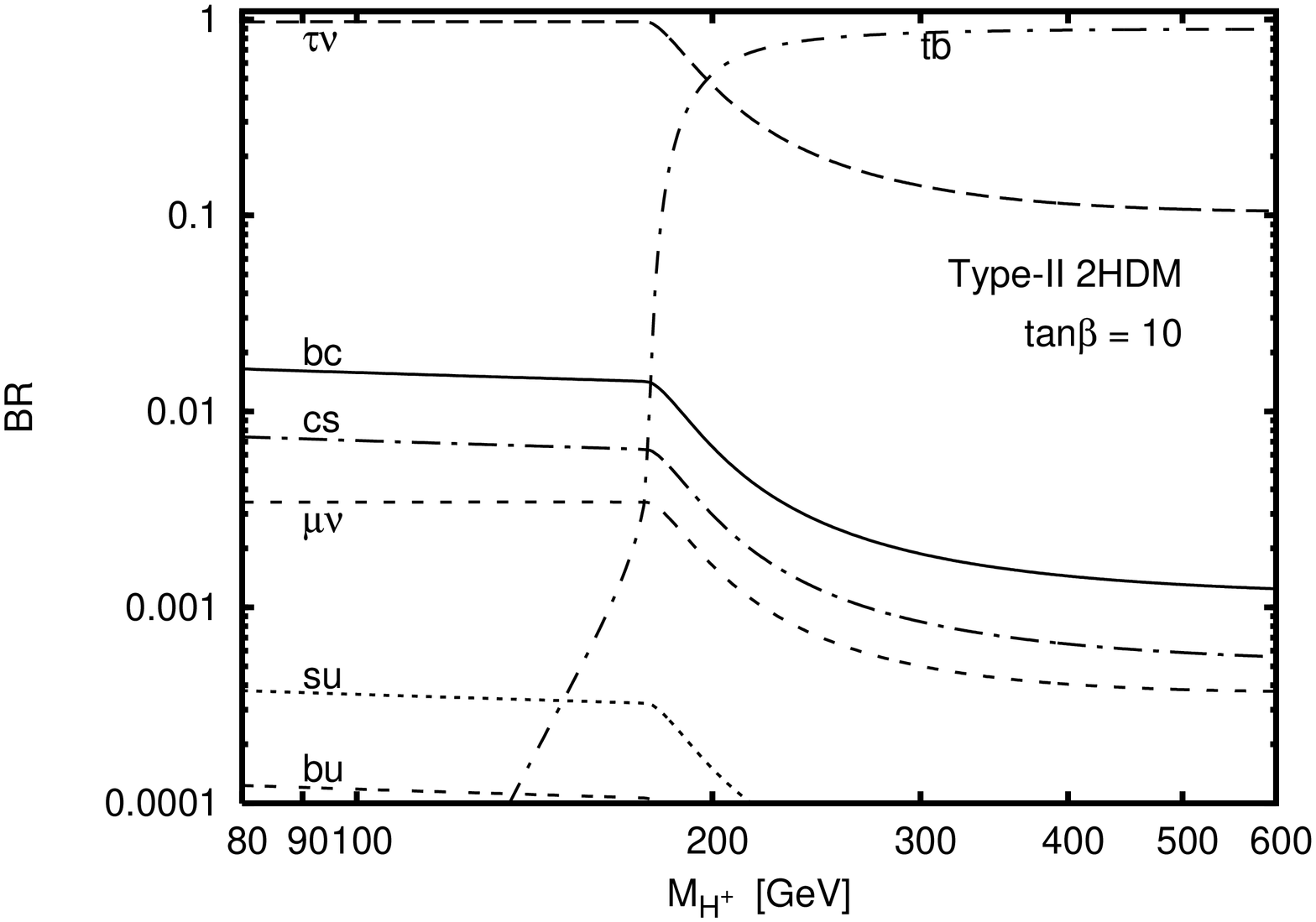}}}
\caption{As in Fig.~\ref{fig:flip5} but for $\tan\beta=10$.}
\label{fig:flip10}
\end{figure}

\begin{figure}
\resizebox{\textwidth}{!}{\rotatebox{270}{\includegraphics{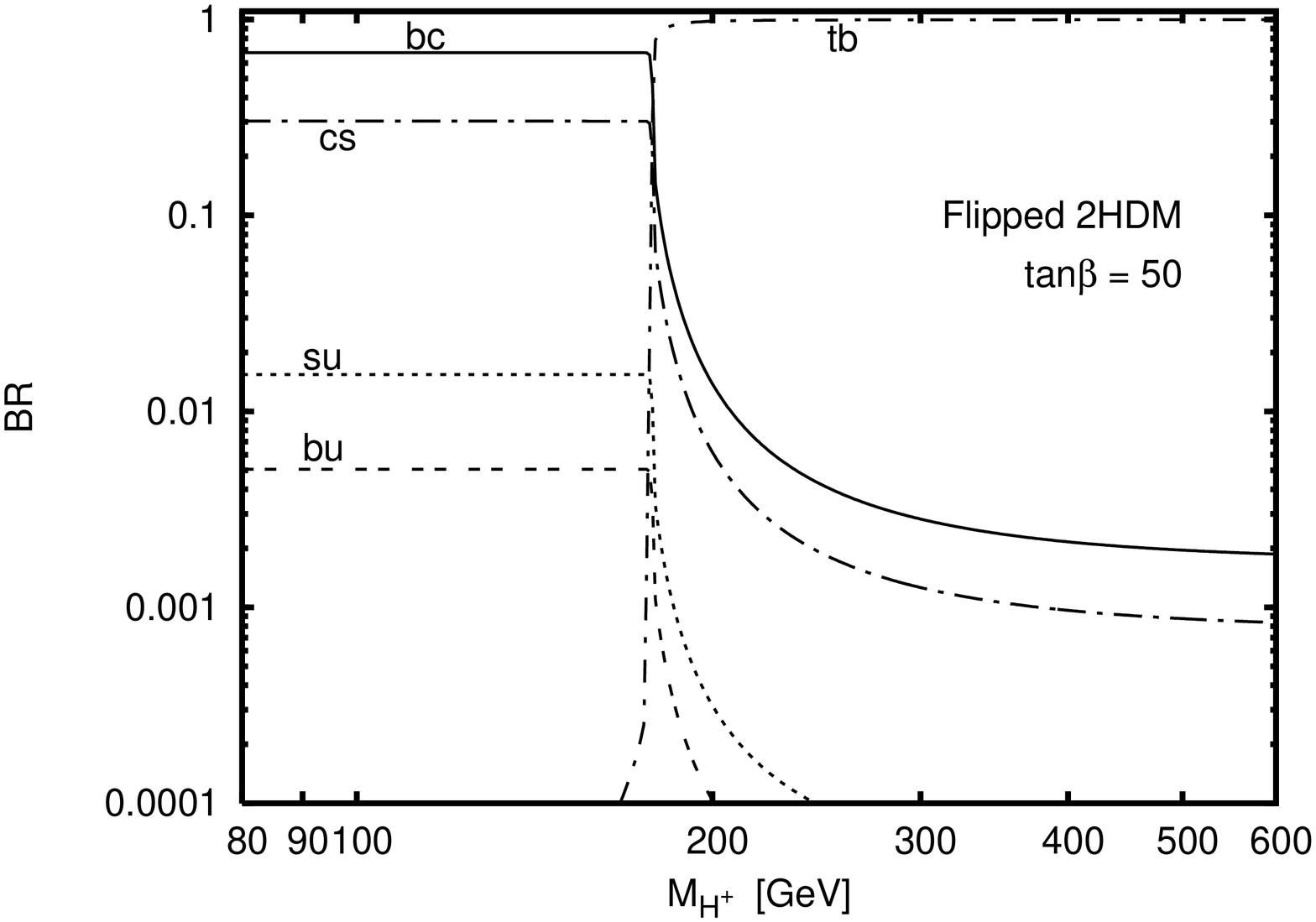}}
\rotatebox{270}{\includegraphics{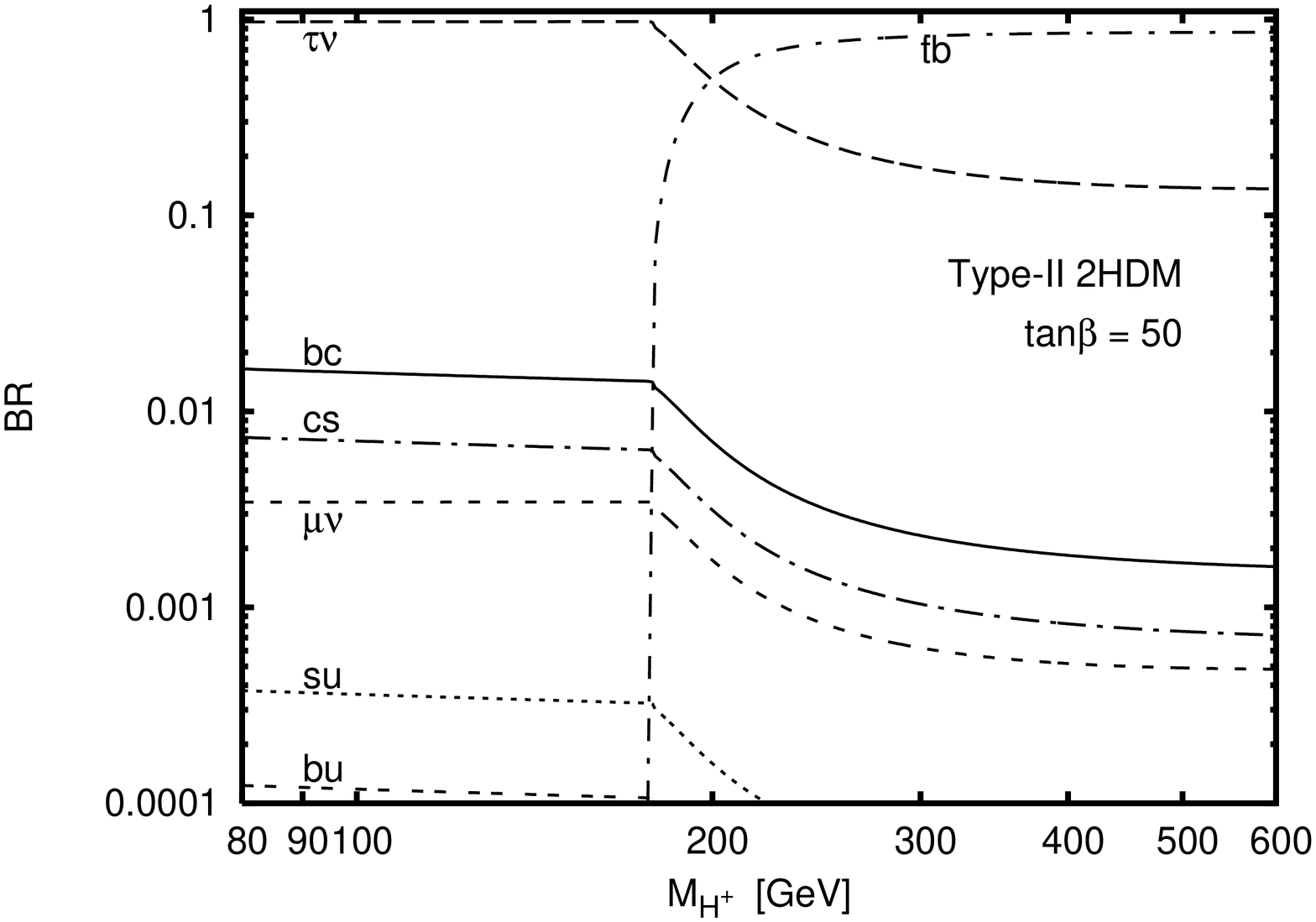}}}
\caption{As in Fig.~\ref{fig:flip5} but for $\tan\beta=50$.}
\label{fig:flip50}
\end{figure}

The branching fractions of the charged Higgs in the flipped model are
shown as a function of $M_{H^+}$ in Figs.~\ref{fig:flip1},
\ref{fig:flip5}, \ref{fig:flip10}, and \ref{fig:flip50} for
$\tan\beta=1$, 5, 10, and 50, respectively.  For comparison, the
corresponding branching fractions in the Type II model are also shown
for $\tan\beta \neq 1$.  The most consistent feature in both models is
the turn-on of decays to $t \bar b$ at the kinematic threshold at
$M_{H^+} \simeq 180$~GeV.  Above this threshold, decays to $t \bar b$
dominate for both the flipped and Type II models, for all values of
$\tan\beta$.  As we will see in Sec.~\ref{sec:indirect}, this high
mass range is favored by the constraint from $b \to s \gamma$.

\begin{figure}
\resizebox{\textwidth}{!}{\rotatebox{270}{\includegraphics{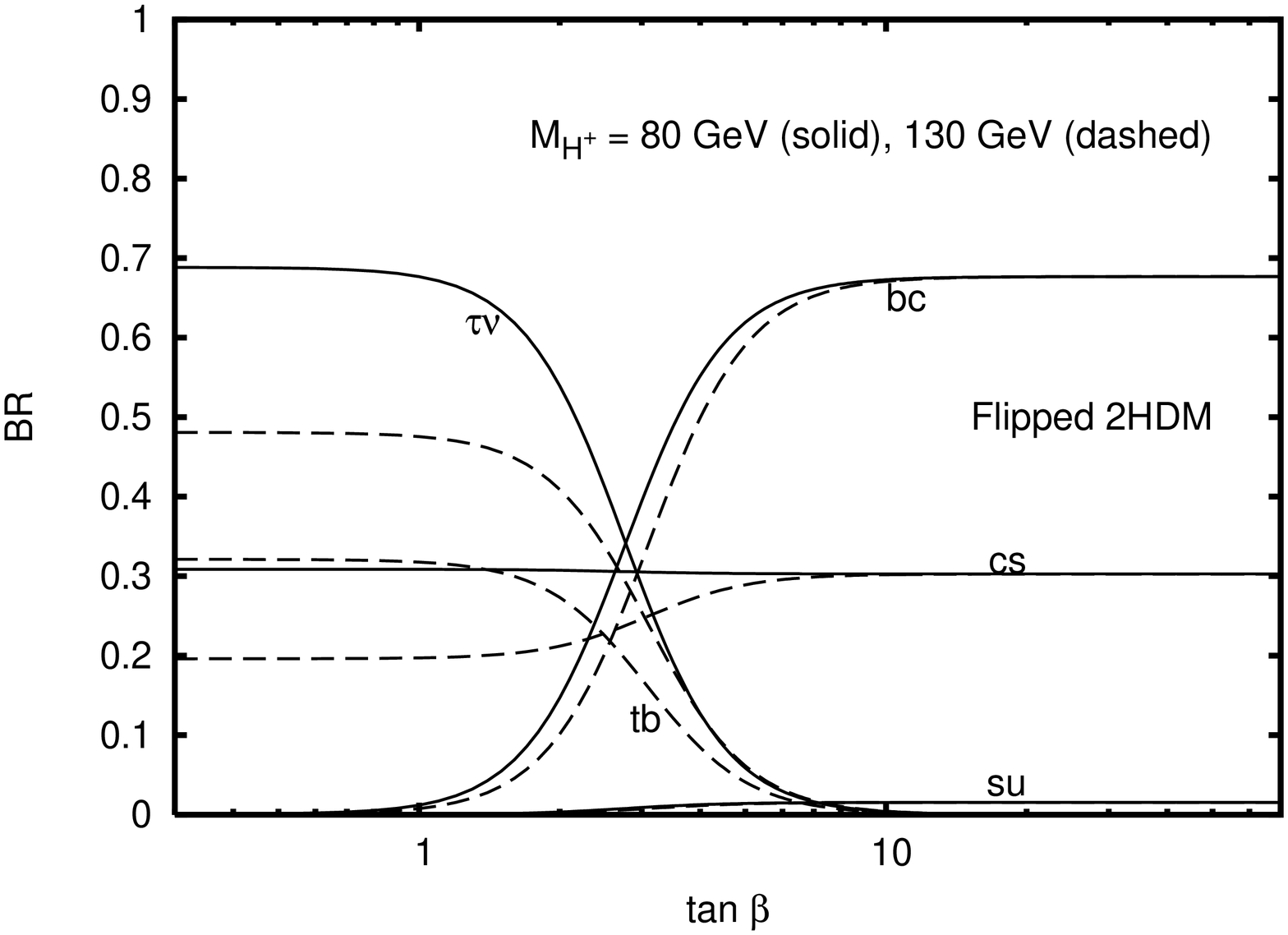}}
\rotatebox{270}{\includegraphics{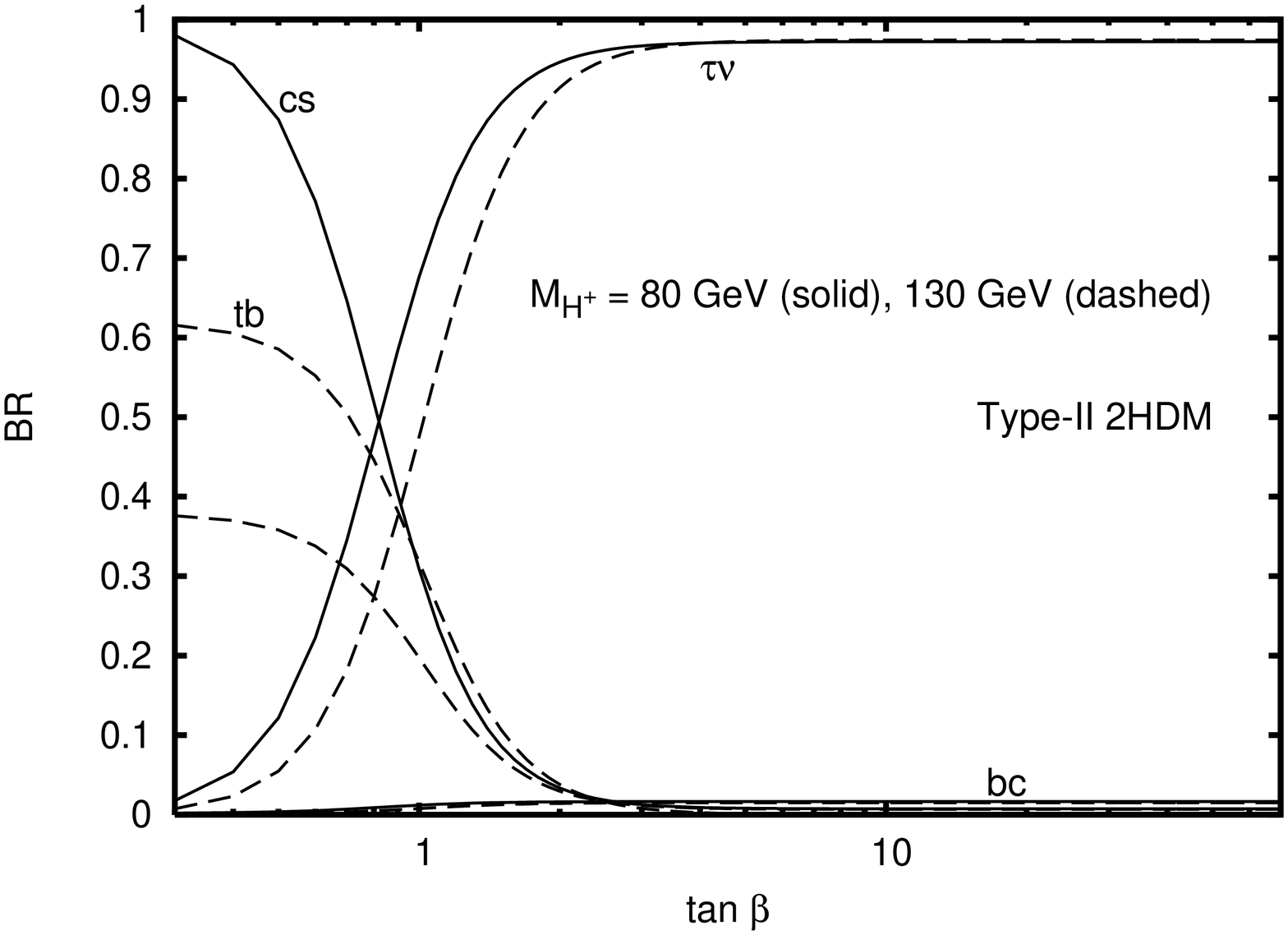}}}
\caption{Charged Higgs branching ratios as a function of $\tan\beta$
for $M_{H^+} = 80$~GeV (solid lines) and 130~GeV (dashed lines) in the
flipped (left) and Type-II (right) 2HDMs.  Note that off-shell decays
to $t \bar b$ appear at low $\tan\beta$ as the charged Higgs mass
increases.}
\label{fig:tanbbrs}
\end{figure}

Below the $t \bar b$ threshold, the effect of the different
$\tan\beta$ dependence of the lepton couplings in the flipped and Type
II models becomes apparent.  We show the branching fractions as a
function of $\tan\beta$ in Fig.~\ref{fig:tanbbrs} for $M_{H^+} = 80$
and 130~GeV, for both the flipped and Type-II models.  For $\tan\beta
= 1$ (Fig.~\ref{fig:flip1}), the branching fractions of $H^+$ in the
flipped model are identical to those in the Type-II model.  As
$\tan\beta$ increases, the branching fraction to $\tau\nu$ is
suppressed in the flipped model due to the $\cot\beta$ dependence of
the lepton Yukawa couplings.  For $\tan\beta = 5$
(Fig.~\ref{fig:flip5}), decays to $\tau\nu$ reach at most $\sim$5\% in
the flipped model, while they dominate below the $t\bar b$ threshold
in the Type II model.  For $\tan\beta = 50$, the branching fraction to
leptons is below $10^{-4}$.  Instead, the dominant decay mode of $H^+$
for $\tan\beta \gtrsim 3$ is into $c \bar b$ with a branching fraction
of about 2/3, followed by $c \bar s$ with a branching fraction of
about 1/3.  The relative strength of these two decays at moderate to
large $\tan\beta$ is controlled by the ratio $V_{cb} m_b / V_{cs} m_s
> 1$ (note that in the flipped model the $c \bar s$ branching fraction
inherits the large uncertainty in the strange quark mass).  The
dominance of the $c \bar b$ mode is in contrast to the Type-II 2HDM;
in that model decays to quarks dominate only at low $\tan\beta
\lesssim 1$ where the charm quark Yukawa coupling contributes
significantly to the rate, leading to ${\rm BR}(H^+ \to c \bar b)/{\rm
BR}(H^+ \to c \bar s) \ll 1$.

\begin{figure}
\resizebox{\textwidth}{!}{\rotatebox{270}{\includegraphics{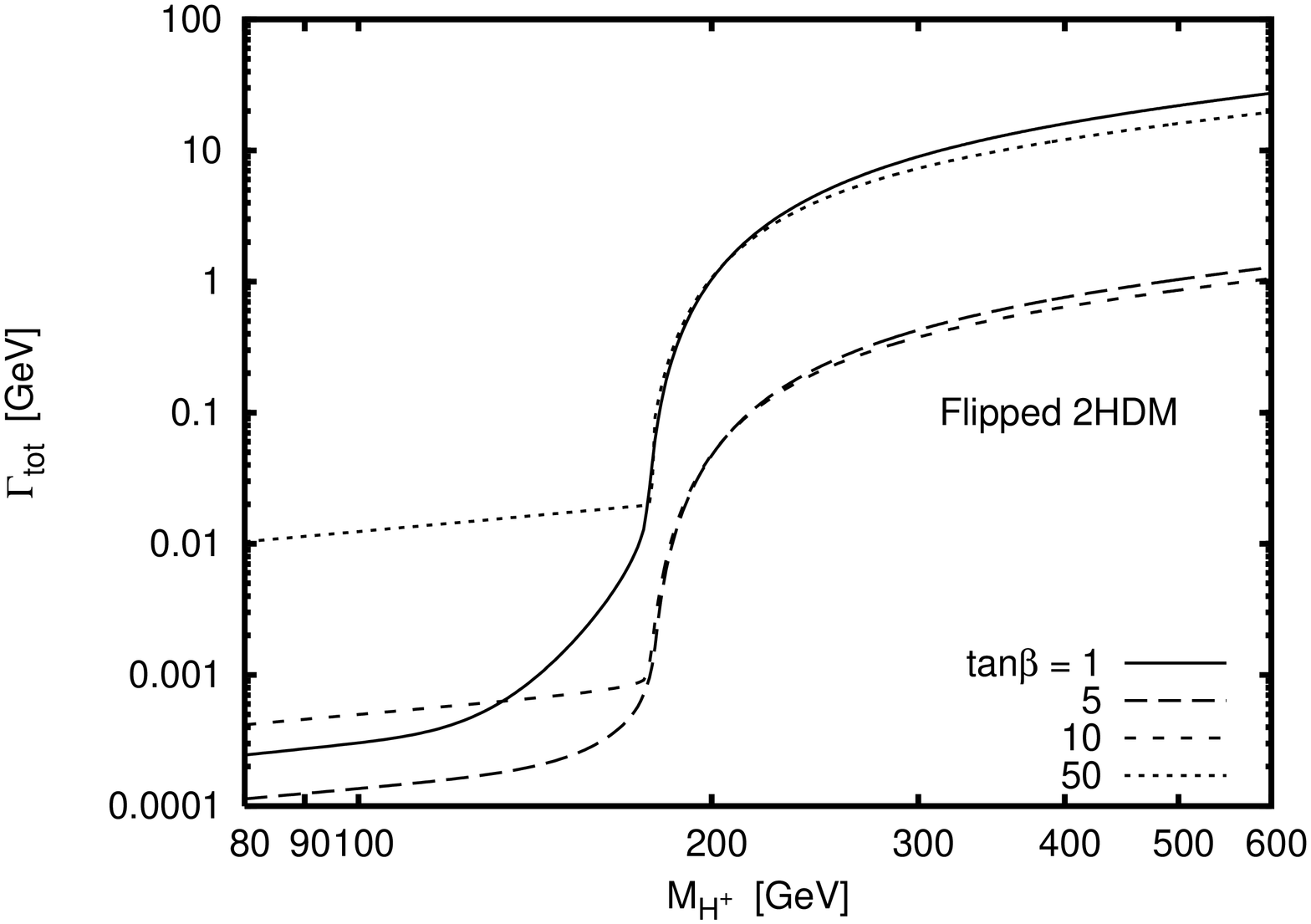}}
\rotatebox{270}{\includegraphics{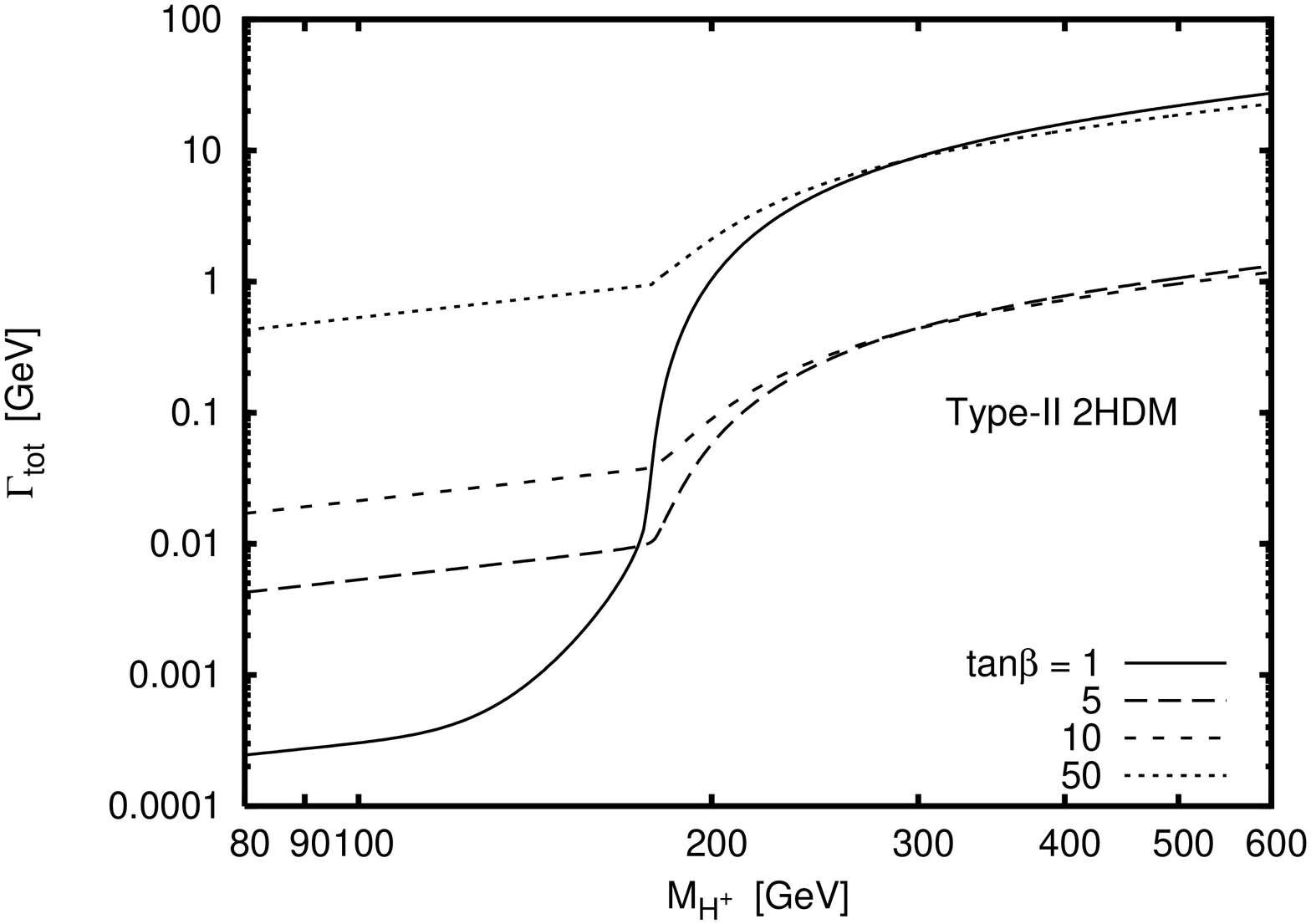}}}
\caption{Total width of the charged Higgs as a function of
$M_{H^+}$, for $\tan\beta = 1$, 5, 10, and 50, in the flipped
(left) and Type-II (right) 2HDMs.}
\label{fig:fliptw}
\end{figure}

The total width of the charged Higgs in the flipped and Type-II models
is shown in Fig.~\ref{fig:fliptw} as a function of $M_{H^+}$ for
various $\tan\beta$ values.  At $\tan\beta = 1$ the Yukawa couplings
are identical in the two models; therefore the total widths are the same.
Above the $tb$ threshold at $M_{H^+} \simeq 180$~GeV the total widths
are again very similar because the quark Yukawa couplings have
identical $\tan\beta$ dependence in the two models.  Below the $tb$
threshold the widths are different in general; in particular, the
charged Higgs in the flipped model is much narrower at large
$\tan\beta$ because the dominant decays are controlled by couplings
proportional to $V_{cb} m_b \tan\beta$ and $V_{cs} m_s \tan\beta$,
which are much smaller than the coupling proportional to $m_{\tau}
\tan\beta$ that controls the dominant decay in the Type II model.

\section{Indirect constraints}
\label{sec:indirect}

Indirect constraints on the charged Higgs in two-doublet models come
from low-energy processes in which an off-shell $H^+$ is exchanged,
either at tree level or at one loop.  Tree-level processes include the
heavy quark decays $B^+ \to \tau^+ \nu$, $D^+ \to \tau^+ \nu$, and $b
\to c \tau \nu$, in which $H^+$ exchange can shift the decay branching
fraction relative to the SM prediction, as well as decays of the
$\tau$ lepton in which the ratio of rates to $\mu \nu \bar \nu$ versus
$e \nu \bar \nu$ provides sensitivity to the mass-dependent lepton
couplings of $H^+$.  The former have been used to constrain the
Type-II
2HDM~\cite{Hou:1992sy,Akeroyd:2008ac,Grossman:1994ax,Grossman:1995yp}
and the latter to constrain the lepton-specific
2HDM~\cite{Aoki:2009ha,Logan:2009uf}.  None of these processes provide
constraints in the flipped model, however---the heavy quark decays
involve a product of the $H^+$ couplings to quarks and to leptons, in
which the $\tan\beta$ dependence cancels in the flipped model, while
the charged Higgs contribution to the $\tau$ decay amplitude is
proportional to $\cot^2\beta$ which is constrained not to become too
large by the perturbativity bound on the top quark Yukawa coupling.

This leaves one-loop processes.  The strongest constraint on $M_{H^+}$
in the flipped model comes from the radiative decay $b \to s \gamma$;
low values of $\tan\beta \lesssim 2$ are also constrained by the mass
difference $\Delta M_{B_d}$ in $B^0$--$\bar B^0$ oscillations.
The $b$ quark fraction in hadronic $Z$ decays, $R_b \equiv \Gamma(Z
\to b \bar b)/\Gamma(Z \to {\rm hadrons})$, provides a weaker constraint
for $\tan\beta \lesssim 1$.  We note however that constraints from
one-loop processes are vulnerable to additional new physics that can
interfere destructively with the charged Higgs contribution,
potentially reopening apparently-excluded parts of parameter space.

\subsection{$b \to s \gamma$}

The strongest indirect constraint on the charged Higgs mass in the
flipped 2HDM comes from the process $b \to s \gamma$.  In the SM, this
process proceeds through one-loop diagrams involving a $W$ boson and
top quark.  In models with a second Higgs doublet there are additional
diagrams in which the $W$ is replaced by a charged Higgs.  Because the
charged Higgs contribution to $b \to s \gamma$ depends only on the
charged Higgs couplings to quarks, the constraints in the flipped 2HDM
are identical to those in the Type-II model, which have been
thoroughly studied.

The measured rate for $b \to s \gamma$ yields a lower bound on
$M_{H^+}$ which is almost independent of $\tan\beta$ for $\tan\beta
\gtrsim 2$; for smaller values of $\tan\beta$ the bound becomes
stronger.  Because the allowed charged Higgs contribution is quite
small, the bound is very sensitive to the experimental central value
and the theoretical calculation for the SM rate.  The current
state-of-the-art theoretical calculation includes QCD corrections at
next-to-next-to-leading order~\cite{Misiak:2006zs} for the SM part and
next-to-leading order for the charged Higgs part.  Using the world
average experimental measurement from Ref.~\cite{Barberio:2006bi},
combining all theoretical and experimental errors in quadrature, and
requiring that BR($b \to s \gamma$) lie within its 95\% (99\%)
confidence level (CL) bound (a one-degree-of-freedom constraint)
yields a lower bound on $M_{H^+}$ of about 295
(230)~GeV~\cite{Misiak:2006zs} for $\tan\beta \gtrsim 2$.  A separate
analysis~\cite{WahabElKaffas:2007xd} treating $M_{H^+}$ and
$\tan\beta$ as two separate degrees of freedom in a chi-squared fit (a
two-degree-of-freedom constraint) found $M_{H^+} \gtrsim 220$~GeV at
95\% CL.

\subsection{$\Delta M_{B_d}$ and $R_b$}

The mass difference $\Delta M_{B_d}$ measured in $B^0$--$\bar B^0$
oscillations arises in the SM from box diagrams involving internal $W$
bosons and top quarks.  In two-doublet models, diagrams in which one
or both of the $W$ bosons is replaced by a charged Higgs give rise to
additional contributions that grow with the top quark Yukawa coupling,
yielding a lower bound on $\tan\beta$ as a function of the charged
Higgs mass~\cite{Geng:1988bq}.  Again, because these contributions
depend only on the charged Higgs coupling to quarks, the constraints
in the flipped 2HDM are identical to those in the Type-II model.
Constraints from this process have been studied recently in
Ref.~\cite{Mahmoudi:2009zx}, which found $\tan\beta \gtrsim 2.0$ (1.3,
0.9) for $M_{H^+} = 100$ (250, 500) GeV at 95\% CL (two-sided
one-degree-of-freedom constraint).

The decay $Z \to b \bar b$ receives corrections in two-Higgs-doublet
models from loops involving $H^+$ and top quarks and from loops
involving neutral Higgs bosons and bottom quarks~\cite{Denner:1991ie}.
The former are important at low $\tan\beta$ when the top quark Yukawa
coupling is enhanced.  The latter are important only at high
$\tan\beta$ and can be neglected for our purposes.  Constraints from
this process were studied recently for the Type II model in
Ref.~\cite{WahabElKaffas:2007xd}, which found that they exclude low
values of $\tan\beta \lesssim 1$ already disfavored by the $\Delta
M_{B_d}$ constraint.

\section{Direct search constraints}
\label{sec:direct}

We now consider constraints from direct charged Higgs searches
performed at LEP and the Tevatron in the context of the Type-II 2HDM,
and show how the resulting limits can be adapted to the flipped model.
While the Higgs masses currently accessible to direct searches are
already disfavored by $b \to s \gamma$, the direct limits cannot be
avoided by introducing additional new physics in the loops.

\subsection{Limits from LEP}

The LEP combined limit on charged Higgs pair production~\cite{LEP2001}
is obtained under the assumption that ${\rm BR}(H^+ \to \tau \nu) +
{\rm BR}(H^+ \to c \bar s) = 1$.  While valid in the Type-II 2HDM,
this assumption does not hold in the flipped model; in particular,
$H^+ \to c \bar b$ is the dominant decay for $\tan\beta \gtrsim 3$ for
charged Higgs masses relevant to the LEP search.  This matters because
the charged Higgs search at DELPHI~\cite{DELPHI} actively selected for
the $c\bar s$ mode, rejecting $b$ jets using the charm mass and decay
multiplicity.  

\begin{figure}
\resizebox{\textwidth}{!}{
\rotatebox{270}{\includegraphics{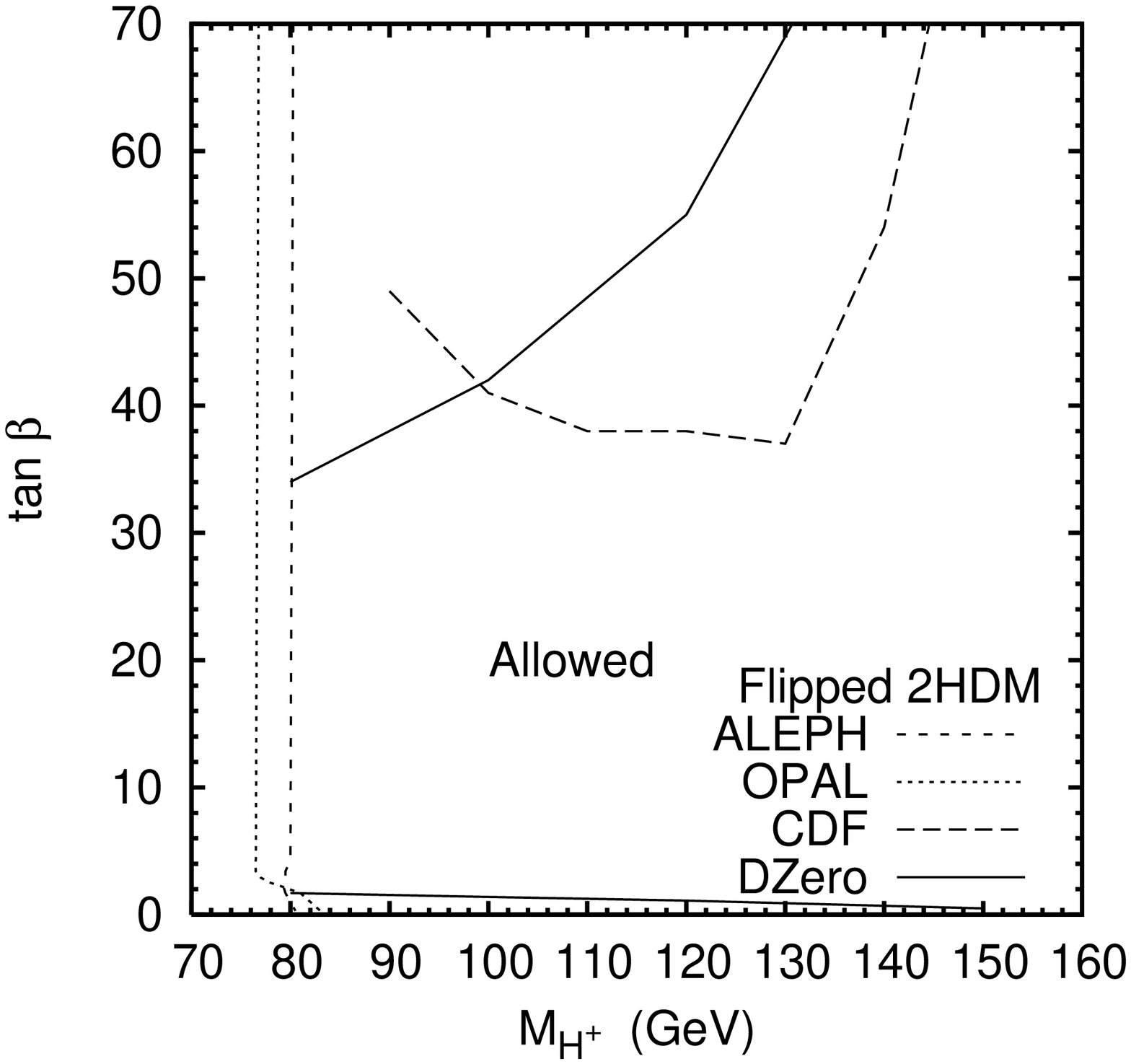}}
\rotatebox{270}{\includegraphics{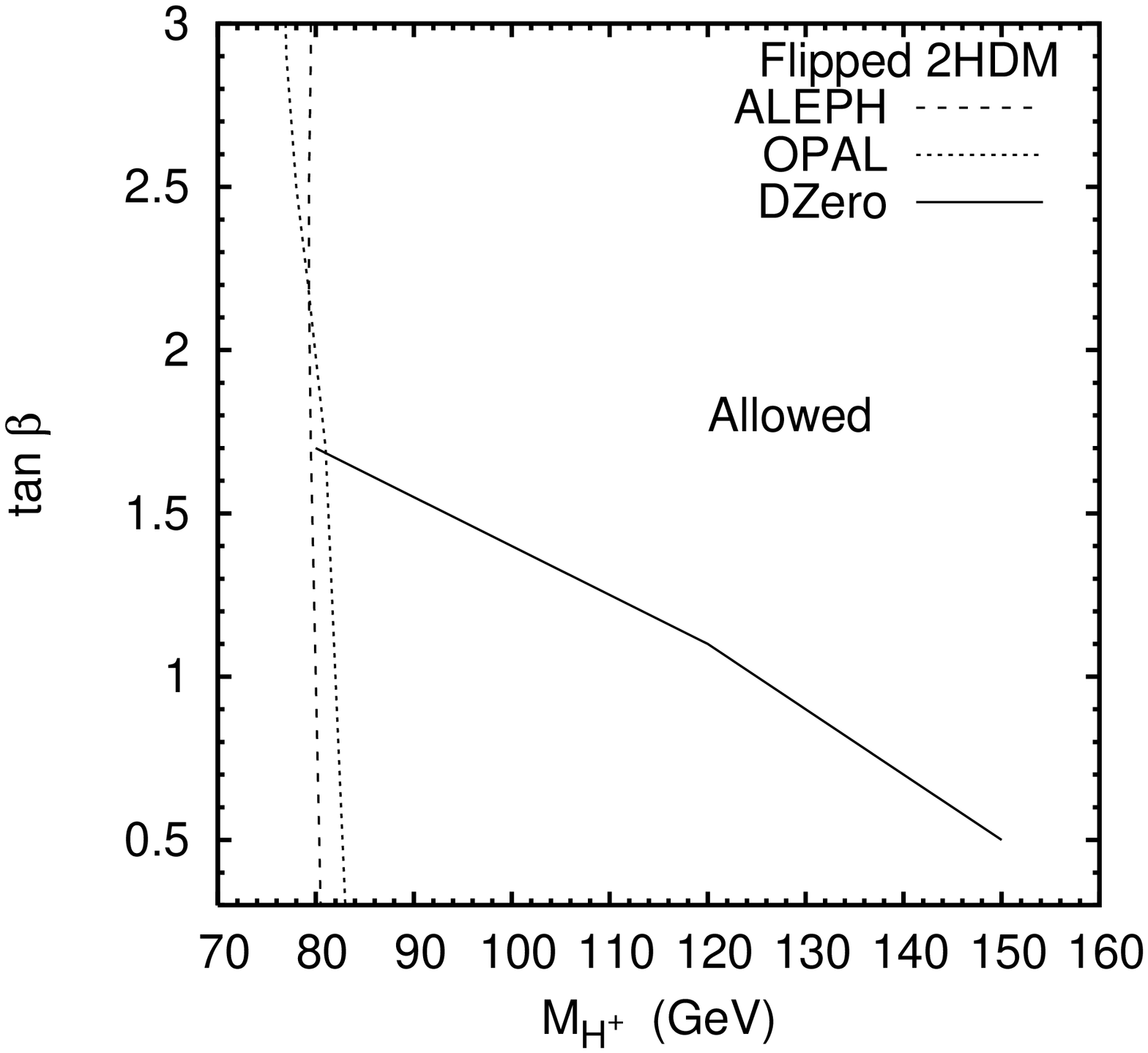}}}
\caption{Direct search limits on the charged Higgs mass and
$\tan\beta$ in the flipped 2HDM.  ALEPH excludes $M_{H^+} < 80$~GeV
over most of the $\tan\beta$ range, with a slightly stronger
constraint from OPAL at low $\tan\beta$ where charged Higgs decays to
$\tau \nu$ become significant.  Large and small values of $\tan\beta$
are excluded by searches at CDF and D{\O} for charged Higgs production
in top quark decay.  The plot on the right shows a close-up view of
the low $\tan\beta$ region.}
\label{fig:directsearch}
\end{figure}

In contrast, the searches at ALEPH~\cite{ALEPH}, L3~\cite{L3}, and
OPAL~\cite{OPAL} were insensitive to the final-state quark flavor and
assumed only that ${\rm BR}(H^+ \to \tau \nu) + {\rm BR}(H^+ \to q
\bar q) = 1$, which is valid in the flipped model.  The most stringent
95\% CL limit over most of the parameter space comes from ALEPH, with
a slightly stronger limit from OPAL at low $\tan\beta$ where the
charged Higgs decay to $\tau \nu$ dominates.  The limits
are~\cite{ALEPH,OPAL},
\begin{equation}
  M_{H^+} \geq \left\{ \begin{array}{cl}
    79.3 \ {\rm GeV} & {\rm overall \ (ALEPH)} \\
    80.4 \ {\rm GeV} & {\rm for \ BR}(H^+ \to q \bar q) = 1 \ {\rm (ALEPH)} \\
    83.0 \ {\rm GeV} & {\rm for \ BR}(H^+ \to \tau \nu) = 0.7 \ {\rm (OPAL)}.
    \end{array} \right.
\end{equation}
Note that for charged Higgs masses around 80~GeV, the branching
fraction to $\tau\nu$ never rises above 0.7 in the flipped model
(see Fig.~\ref{fig:tanbbrs}).  These limits are shown in
Fig.~\ref{fig:directsearch} along with constraints from the Tevatron,
which we discuss next.

\subsection{Limits from the Tevatron}

The Tevatron experiments CDF and D{\O} have searched for charged Higgs
bosons produced in top quark decay (i.e., for $M_{H^+} < m_t - m_b$)
with $H^+$ decaying to $c \bar s$ or
$\tau\nu$~\cite{Aaltonen:2009ke,Abazov:2009zh}.  Limits from these
searches on $M_{H^+}$ and $\tan\beta$ have been presented in the
context of the Type-II 2HDM.  While the partial width for $t \to H^+
b$ is the same in the flipped and Type-II models, the decay branching
fractions for $H^+$ depend differently on $\tan\beta$, leading to
different limits in the flipped model which we derive here.

CDF~\cite{Aaltonen:2009ke} searched for charged Higgs bosons decaying
to $c \bar s$ in $t \bar t$ events by looking for a second resonance
(in addition to the resonance from $W \to jj$) in the invariant mass
distribution of two jets in the lepton-plus-jets $t \bar t$ sample.
Nonobservation of a second resonance allowed CDF to set a limit on
BR($t \to H^+ b$) as a function of $M_{H^+}$, assuming that BR($H^+
\to c \bar s) = 1$.  Limits on BR($t \to H^+ b)$ are in the range
0.08--0.32 for $M_{H^+}$ between 90 and 150~GeV.  No limit is set for
$M_{H^+}$ between 70 and 90 GeV due to overwhelming background from
the $W$ mass peak.

In the flipped 2HDM, BR($H^+ \to q \bar q^{\prime}) \simeq 1$ at large
$\tan\beta$, with decays predominantly to $c \bar b$.  The only
difference in the reconstruction of the charged Higgs in this case
should be a minor degradation of the energy resolution of the $H^+$
mass peak due to the poorer energy resolution of bottom quark jets;
we expect that this should not significantly reduce the sensitivity.
We apply the CDF constraint to the flipped 2HDM at large $\tan\beta$
by computing BR($t \to H^+ b)$ at tree level as a function of
$M_{H^+}$ and $\tan\beta$ and taking BR($H^+ \to q \bar q^{\prime}) =
1$ (see the Appendix for details).  This yields an upper bound on
$\tan\beta$ in the range of 40--50 for $M_{H^+}$ between 90 and 140
GeV, as shown in Fig.~\ref{fig:directsearch}.

D{\O}~\cite{Abazov:2009zh} searched for charged Higgs production in
top quark decays through a fit of the event rates in several
single-lepton and two-lepton top quark channels, using the SM
prediction for the $t \bar t$ cross section and known $W$ decay
branching ratios.  Charged Higgs production with decays to $q \bar
q^{\prime}$ would lead to an overall decrease in the rates for
leptonic channels, while charged Higgs production with decays to $\tau
\nu$ would increase rates in the hadronic tau channel while decreasing
other channels.  This allows an upper limit on BR($t \to H^+ b$) of
about 0.2 to be set largely independent of the charged Higgs mass.
Limits were obtained as a function of BR($H^+ \to q \bar q^{\prime})$
under the assumption that the charged Higgs decays only to $q \bar
q^{\prime}$ and $\tau \nu$.

We apply the D{\O} limits to the flipped 2HDM as follows.  At large
$\tan\beta$ we take BR($H^+ \to q \bar q^{\prime}) = 1$ and compute
BR($t \to H^+ b$) at tree level as a function of $M_{H^+}$ and
$\tan\beta$.  The resulting constraint is weaker than that from the
CDF direct search except for $M_{H^+} \simeq 80$--100~GeV, where the
CDF search is degraded by $W$ background; the resulting upper limit on
$\tan\beta$ is 34--42.  At low $\tan\beta$ we compute the
branching fraction of $H^+ \to q \bar q^{\prime}$ using our modified
version of {\tt HDECAY} and translate the appropriate upper limit on
BR($t \to H^+ b$) into a lower limit on $\tan\beta$.  The resulting
limit ranges from 1.7 for $M_{H^+} = 80$~GeV to 0.5 at $M_{H^+} =
150$~GeV.  Results for both large and small $\tan\beta$ are shown in
Fig.~\ref{fig:directsearch}.

\section{LHC search prospects}
\label{sec:lhc}

We now review the existing LHC charged Higgs search studies and
discuss their implications for the flipped 2HDM.  Existing ATLAS and
CMS studies focus on charged Higgs production via the $H^+ \bar t b$
coupling: in particular, $t \bar t$ production with $t \to H^+ b$ and
$t H^{\pm}$ associated production.  Because the quark coupling
structure in the flipped model is identical to that in the usual
Type-II 2HDM, the charged Higgs production cross sections in these
modes are identical in the two models.  Likewise, for $M_{H^+} \gtrsim
m_t + m_b$, decays of $H^+$ are predominantly to $t \bar b$ in either
model.  The flipped model phenomenology differs from that of the
Type-II model for charged Higgs masses below the $t \bar b$ threshold,
where decays of $H^+ \to q \bar q^{\prime}$ dominate for $\tan\beta$
greater than a few, and above the $t \bar b$ threshold insofar as 
decays to $\tau \nu$ are always completely negligible.

For completeness, we also discuss $H^+H^-$ pair production through
electroweak and Yukawa interactions and $H^+W^-$ associated production
through Yukawa interactions.

\subsection{$t \to H^+ b$}

ATLAS~\cite{hep-ex/0901.0512} and CMS~\cite{Baarmand:2006dm} have
studied $t \bar t$ production with one top quark decaying to a charged
Higgs, $t \to H^+ b$; however, all such studies to date have assumed
the charged Higgs decay $H^+ \to \tau \nu$.  While this mode provides
good parameter space coverage in the Type-II 2HDM, its range of
applicability in the flipped 2HDM is already largely constrained by the
D{\O} search discussed in the previous section.

Instead, the most relevant search channel in the flipped model would
involve the decay $H^+ \to q \bar q^{\prime}$, as considered already
at the Tevatron~\cite{Aaltonen:2009ke,Abazov:2009zh}.  We note that
the dominant hadronic decay is $H^+ \to c \bar b$; $b$ tagging may
thus enhance the signal while at the same time providing evidence for
the flipped model Yukawa structure.

\subsection{$H^{\pm}t$ associated production}

The charged Higgs boson can be produced in association with a top
quark through bottom-gluon fusion, $gb \to t H^-$, and through
gluon-gluon fusion, $gg \to \bar b t H^-$.  The inclusive cross
section has been computed to next-to-leading order in
QCD~\cite{Plehn:2002vy}.  The cross section grows at large $\tan\beta$
proportional to $(m_b \tan\beta)^2$; for $M_{H^+} = 250$~GeV and
$\tan\beta = 30$ it reaches about 500~fb~\cite{Plehn:2002vy}.  Above
the $t\bar b$ threshold, the charged Higgs in the flipped 2HDM decays
almost exclusively to $t \bar b$.  Studies of this channel done in the
context of the Type-II model can thus be directly applied to the
flipped model.

ATLAS~\cite{hep-ex/0901.0512} and CMS~\cite{CMStHtb} have studied the
channel $tH^-$, $H^- \to \bar t b$ in the context of the MSSM.  The
charged Higgs production cross section in these studies is identical
to that in the flipped model, up to $\tan\beta$-enhanced corrections
to the bottom quark Yukawa coupling from loops involving
supersymmetric particles which can affect the $tH^-$ associated
production cross section at large $\tan\beta$~\cite{Plehn:2002vy}.  We
ignore these corrections here.  The major background comes from $t
\bar t$ plus jets.  Unfortunately, both experiments conclude that
given the systematic uncertainty on the background, this channel
yields no discovery potential for $\tan\beta$ values below 100 with 30
fb$^{-1}$ of integrated luminosity.  ATLAS combines this $t H^-$, $H^-
\to \bar t b$ channel with $t H^-$, $H^- \to \tau \nu$ in the MSSM to
present combined discovery reach contours at large $\tan\beta$, but
the $H^- \to \bar t b$ contribution improves the reach only
marginally~\cite{hep-ex/0901.0512}.  We emphasize that while BR($H^-
\to \tau \nu) \sim 10\%$ above the $tb$ threshold in the MSSM at large
$\tan\beta$, in the flipped model this decay mode is completely
negligible at large $\tan\beta$.

\subsection{$H^+H^-$ pair production}

Charged Higgs bosons can be pair produced via electroweak interactions
(Drell-Yan $q \bar q \to \gamma^*, Z^* \to H^+H^-$ and weak boson
fusion $qq \to qq H^+H^-$) and via Yukawa interactions (bottom quark
fusion through $t$-channel top quark exchange, $b \bar b \to H^+H^-$,
and gluon fusion through a one-loop process involving bottom and top
quarks in the loop, $gg \to H^+H^-$; diagrams involving $s$-channel
exchange of the neutral scalars $h^0$ and $H^0$ also contribute).
Cross sections in these modes in the flipped model are identical to
those in the Type-II 2HDM; representative cross sections are quoted in
Table~\ref{tab:crosssec} for $M_{H^+} = 200$~GeV.

\begin{table}
\begin{tabular}{lrcccc}
\hline\hline
\multicolumn{6}{c}{Cross sections in fb for $M_{H^+} = 200$ GeV} \\
Process & $\tan\beta = 10$ & ~~~~20~~ & ~~30~~ & ~~50~~~~ & Source\\
\hline
$q \bar q \to H^+H^-$ & \multicolumn{4}{c}{26 for all $\tan\beta$} 
   & \cite{Alves-Plehn} \\
$b \bar b \to H^+H^-$ & 0.1 & 0.1 & 0.5 & 6 & \cite{Alves-Plehn} \\
$gg \to H^+H^-$ & 0.2 & 2 & 10 & 70 & \cite{Alves-Plehn} \\
$qq \to qqH^+H^-$ & \multicolumn{4}{c}{17 for all $\tan\beta$} 
    & \cite{Moretti:2001pp,Logan:2009uf} \\
\hline\hline
\end{tabular}
\caption{Charged Higgs pair production cross sections in various modes
at the LHC.}
\label{tab:crosssec}
\end{table}

The Drell-Yan process and the two Yukawa processes were studied in
detail at NLO in Ref.~\cite{Alves-Plehn}.  At low to moderate
$\tan\beta$ values, the Drell-Yan process dominates with a cross
section of about 26~fb for $M_{H^+} = 200$~GeV.  The Yukawa processes
are enhanced at large $\tan\beta$, with cross sections growing like
$(m_b \tan\beta)^4$; the gluon fusion process dominates at large
$\tan\beta$ values, reaching $\sim$70~fb at $\tan\beta = 50$.

The weak boson fusion process was studied in detail in
Ref.~\cite{Moretti:2001pp}.  It does not depend significantly on
$\tan\beta$; for $M_{H^+} = 200$~GeV we found a tree-level cross
section at the LHC of 17~fb~\cite{Logan:2009uf}.  While this cross
section is rather small, it falls more slowly with increasing
$M_{H^+}$ than the $s$-channel processes and the two forward jets
may provide useful kinematic handles against QCD backgrounds.

For charged Higgs masses below the $tb$ threshold, the dominant decay
for $\tan\beta$ larger than a few is $H^+ \to c \bar b$, leading to a
signal in these channels of $c \bar b \bar c b$.  We expect that this
will be completely swamped by four-jet QCD backgrounds.  Charged Higgs
production in top quark decay has already proven to be a more
sensitive channel at the Tevatron and we expect that this will remain
true at the LHC.

For charged Higgs masses above the $tb$ threshold, the signal in any
of these channels will be $H^+H^- \to t \bar b \bar t b$.  We expect
that backgrounds will be dominated by $t \bar t$ just as in the case
of the $t H^-$, $H^- \to \bar t b$ channel discussed in the previous
section; at the same time the signal cross section is at least an
order of magnitude smaller than for $t H^-$.  Unless the forward jets
in the weak boson fusion process can be used to severely reduce the
background, we thus expect that these channels will be even less
sensitive than $t H^-$ associated production.

\subsection{$H^{\pm}W^{\mp}$ associated production}

Finally, we consider associated production of a charged Higgs boson
with a $W^{\pm}$, initiated by $b \bar b$ fusion at tree level or
gluon fusion at one loop.  These processes can proceed directly via
$H^+$ couplings to an initial or internal fermion, as well as via
$H^+W^-$ coupling to an $s$-channel neutral Higgs boson ($h^0$, $H^0$,
or $A^0$).  The leading-order cross sections were computed in
Refs.~\cite{Dicus:1989vf,BarrientosBendezu:1998gd} for the MSSM Higgs
sector.  Squark loop contributions to the gluon-fusion process
in the MSSM were computed in Refs.~\cite{Brein:2000cv}, and one-loop
electroweak and QCD corrections to the $b \bar b$ fusion process
were considered in Refs.~\cite{Yang:2000yt} and \cite{Hollik:2001hy}, 
respectively.
For $M_{H^+} = 200$~GeV, Ref.~\cite{BarrientosBendezu:1998gd} found
\begin{equation}
  \left. \begin{array}{r}
   \sigma(b \bar b \to H^{\pm}W^{\mp}) = 200\ (25,\ 300)\ {\rm fb} \\
   \sigma(gg \to H^{\pm}W^{\mp}) = 80\ (5,\ 5)\ {\rm fb} 
   \end{array} \right\} \qquad
         {\rm for}\ \tan\beta = 1.5\ (6,\ 30).
\end{equation}
The $b \bar b$ fusion process is dominant, in part because of strong
destructive interference between quark box and triangle diagrams in the
gluon fusion process~\cite{BarrientosBendezu:1998gd}.
  
The cross sections in the flipped 2HDM are the same as those
considered in Ref.~\cite{BarrientosBendezu:1998gd} for the same Higgs
spectrum and mixing angles, except for the absence of supersymmetric
$\tan\beta$-enhanced one-loop corrections to the bottom quark Yukawa
coupling.  However, the MSSM imposes stringent relations among
the neutral and charged Higgs masses which are absent in a generic
2HDM.  Reference~\cite{Asakawa:2005nx} studied $H^{\pm}W^{\mp}$
production at the LHC in a generic 2HDM and found that, compared to
the MSSM case, the cross section could be enhanced by one to two
orders of magnitude in some parts of parameter space (allowed by all
theoretical and experimental constraints) due to the lifting of
destructive interference or resonant production of a heavier neutral
Higgs boson followed by decay to $H^{\pm}W^{\mp}$.

Reference~\cite{Moretti:1998xq} studied the LHC search prospects in
this channel with $W^{\mp} \to \ell \nu$ and $H^{\pm} \to tb$.  In the
context of an MSSM-like Higgs sector, the signal appears to be
completely swamped by the large $t\bar t$ background; however, in
favorable non-MSSM-like regions of 2HDM parameter space the cross
section could be enhanced to an observable level~\cite{Asakawa:2005nx}.

\section{Conclusions}
\label{sec:conclusions}

Experimental evidence for a charged Higgs boson would constitute
conclusive proof of physics beyond the SM and shed light on the
mechanism of electroweak symmetry breaking.  The simplest extensions of
the SM that contain a charged Higgs are two-Higgs-doublet models.
There are four possible assignments of the fermion couplings in
two-Higgs-doublet models that naturally avoid tree-level
flavor-changing neutral Higgs interactions.  The phenomenology and LHC
search prospects for the charged Higgs boson in the Type-I and -II
models have been studied extensively in the literature; recent papers
have done the same for the lepton-specific 2HDM.  We complete the
picture here by studying the charged Higgs of the flipped 2HDM, in
which one doublet gives mass to up-type quarks and charged leptons and
the other gives mass to down-type quarks.

The phenomenology of the charged Higgs is controlled by the structure
of the Yukawa couplings to quarks and leptons, the charged Higgs mass,
and the parameter $\tan\beta$.  In the flipped model, the charged
Higgs couplings to quarks have the same $\tan\beta$ dependence as in
the usual Type-II model.  This allowed us to carry over directly from
the Type-II model those indirect constraints and charged Higgs
production cross sections that depend only on quark couplings.  In
contrast, the charged Higgs couplings to leptons have the opposite
$\tan\beta$ dependence in the flipped model than in the Type-II model.
This eliminates indirect constraints from (semi-)leptonic meson decays
and leads to a different pattern of charged Higgs decays below the
$tb$ threshold compared to the Type-II model.

We used these features to extract the existing indirect and direct
constraints on the charged Higgs in the flipped model and to evaluate
the search prospects at the LHC.  The strongest indirect constraint
comes from $b \to s \gamma$, which yields $M_{H^+} \gtrsim
220$--300~GeV; $B^0$--$\bar B^0$ mixing also constrains $\tan\beta
\gtrsim 2.0$ (1.3, 0.9) for $M_{H^+} = 100$ (200, 300)~GeV.  Because
indirect constraints are vulnerable to contributions from other
unknown new physics, we also considered direct search constraints on
the charged Higgs from LEP and the Tevatron.  Taking account of the
different decay patterns in the flipped model compared to the Type-II
model, we found that the LEP experiments yield $M_{H^+} \gtrsim 80$
GeV while Tevatron searches for $t \to H^+ b$ put an upper bound on
$\tan\beta$ of about 40 for $M_{H^+} \lesssim 130$~GeV and exclude
very small $\tan\beta$ values below about 0.5--1.5.

It should be possible at the LHC to extend the Tevatron's sensitivity
in $t \to H^+ b$ to more moderate $\tan\beta$ values, but existing
studies consider only the $H^+ \to \tau \nu$ channel, which would
provide sensitivity in the flipped model only at low $\tan\beta$.  We
expect that the LHC studies could be fruitfully extended to include
$H^+ \to q \bar q^{\prime}$ as in the Tevatron searches.

For charged Higgs masses above the $tb$ threshold, the only
potentially promising discovery channel appears to be $tH^-$
associated production with $H^- \to tb$.  Signal rates in the flipped
and Type-II models are essentially identical.  Unfortunately the most
up-to-date ATLAS and CMS studies of this channel with 30~fb$^{-1}$
conclude that the signal is obscured by the systematic uncertainty in
the $t \bar t$ background.  Charged Higgs searches thus remain a major
challenge for the LHC.


\begin{acknowledgments}
This work was supported by the Natural Sciences and Engineering
Research Council of Canada.  
\end{acknowledgments}


\appendix

\section{Charged Higgs production in top quark decay}

We compute the top quark branching fractions as follows.  
Neglecting the bottom quark mass, the SM partial width for $t \to W^+ b$
is given at tree level by,
\begin{equation}
  \Gamma(t \to W^+ b) = \frac{G_F m_t}{8 \sqrt{2} \pi}
  \left[m_t^2 + 2 M_W^2 \right]
  \left[1 - \frac{M_W^2}{m_t^2} \right]^2.
\end{equation}
Taking $m_t = 171.2$~GeV and $M_W = 80.398$~GeV~\cite{PDG} yields
$\Gamma(t \to W^+ b) = 1.44$~GeV.

In the flipped and Type-II models, the partial width for $t \to H^+ b$
is given at tree level by,
\begin{equation}
   \Gamma(t \to H^+ b) = \frac{G_F m_t}{8\sqrt{2}\pi} 
   \left[m_t^2 \cot^2\beta + m_b^2 \tan^2\beta \right]
   \left[1 - \frac{M_{H^+}^2}{m_t^2} \right]^2,
\end{equation}
where we again neglect the final-state bottom quark mass in the
kinematics, but keep $m_b$ where it enters the charged Higgs Yukawa
coupling.  In particular, the decay width becomes large at both low
and high $\tan\beta$.

At high $\tan\beta$ the $H^+ \bar t b$ coupling is dominated by the
term proportional to $m_b \tan\beta$.  This coupling receives large
QCD corrections, which can be mostly accounted for by using the
running bottom quark mass evaluated at the energy scale of the
process, which we take to be $M_{H^+}$.  We use the one-loop
expression for the running bottom quark mass,
\begin{equation}
  \bar m_b(M_{H^+}) = \left[ \frac{\alpha_s(M_{H^+})}{\alpha_s(M_b)} 
    \right]^{4/b_0} \bar m_b(M_b),
\end{equation}
where $b_0 = 11 - 2 n_f/3$ and $n_f = 5$ is the number of active quark
flavors in the energy range of interest.  Here $\bar m_b(M_b) = 4.20
\pm 0.04$~GeV~\cite{Buchmuller:2005zv} is the running bottom quark
mass in the modified minimal subtraction scheme evaluated at the
bottom quark mass scale, and the running strong coupling $\alpha_s$ is
given at one loop by
\begin{equation}
  \alpha_s(Q) = \frac{\alpha_s(M_Z)}
	{1 + (b_0 \alpha_s(M_Z)/2 \pi) \ln (Q/M_Z)}.
\end{equation}
We use $\alpha_s(M_Z) = 0.1176$~\cite{PDG}.


\end{document}